\documentclass[10pt,letterpaper,conference]{IEEEtran}

\usepackage[english]{babel}

\usepackage[latin1]{inputenc}

\usepackage{ae}

\usepackage{amsmath}

\usepackage{amssymb}

\usepackage{exscale}

\usepackage{subfigure}

\usepackage{graphicx}

\usepackage{psfrag}

\usepackage{calc}

\usepackage{url}

\usepackage{layout}

\usepackage{ifthen}

\usepackage{amsfonts}

\usepackage{amsmath}

\usepackage{amssymb}

\usepackage{exscale}

\usepackage{xspace}

\usepackage{dsfont}

\providecommand*{\MathLib}{UNDEFINED}
\ifthenelse{\equal{\MathLib}{UNDEFINED}}{
\renewcommand*{\MathLib}{DEFINED}

\providecommand*{\UseNewTheorem}{ANYTHING}

\providecommand*{\TheoremLayout}{LEFT-HAND-NUMBERS-LAYOUT}

\providecommand*{\thechapter}{UNDEFINED}

\providecommand*{\Xmath}[1]{\ensuremath{{#1}}\xspace}

\ifthenelse{\equal{\UseNewTheorem}{TRUE}}{%

\usepackage{amsthm}

  \ifthenelse{\equal{\thechapter}{UNDEFINED}}{%
    \providecommand*{\ResetCounter}{section}
    
  }{%
    \providecommand*{\ResetCounter}{chapter}
    
  }
  
  \iflanguage{english}{%
    \newcommand*{\TheoremName}{Theorem}
    \newcommand*{\LemmaName}{Lemma}
    \newcommand*{\CorollaryName}{Corollary}
    \newcommand*{\DefinitionName}{Definition}
    \newcommand*{\RemarkName}{Remark}
    \newcommand*{\ExampleName}{Example}
    \newcommand*{\ExerciseName}{Exercise}
    \newcommand*{\ModelName}{Model}
    \newcommand*{\NotationName}{Notation}
    \newcommand*{\CaseName}{Case}
    \newcommand*{\QuestionName}{Question}
    \newcommand*{\AssumptionName}{Assumption}
  }{}
  
  \iflanguage{german}{%
    \newcommand*{\TheoremName}{Satz}
    \newcommand*{\LemmaName}{Lemma}
    \newcommand*{\CorollaryName}{Folgerung}
    \newcommand*{\DefinitionName}{Definition}
    \newcommand*{\RemarkName}{Bemerkung}
    \newcommand*{\ExampleName}{Beispiel}
    \newcommand*{\ExerciseName}{Aufgabe}
    \newcommand*{\ModelName}{Modell}
    \newcommand*{\NotationName}{Bezeichnung}
    \newcommand*{\CaseName}{Fall}
    \newcommand*{\QuestionName}{Frage}
    \newcommand*{\AssumptionName}{Annahme}
  }{}

  \ifthenelse{\equal{\TheoremLayout}{LEFT-HAND-NUMBERS-LAYOUT}}{%
  
    \swapnumbers
  
    \theoremstyle{plain}
    \newtheorem{theorem}{\TheoremName}[\ResetCounter]
    \newtheorem*{theorem*}{\TheoremName}
    \newtheorem{lemma}[theorem]{\LemmaName}
    \newtheorem*{lemma*}{\LemmaName}
    \newtheorem{corollary}[theorem]{\CorollaryName}
    \newtheorem*{corollary*}{\CorollaryName}
  
    \theoremstyle{definition}
    \newtheorem{definition}[theorem]{\DefinitionName}
    \newtheorem*{definition*}{\DefinitionName}
    \newtheorem{remark}[theorem]{\RemarkName}
    \newtheorem*{remark*}{\RemarkName}
    \newtheorem{example}[theorem]{\ExampleName}
    \newtheorem*{example*}{\ExampleName}
    \newtheorem{exercise}[theorem]{\ExerciseName}
    \newtheorem*{exercise*}{\ExerciseName}
    \newtheorem{model}[theorem]{\ModelName}
    \newtheorem*{model*}{\ModelName}
    \newtheorem{notation}[theorem]{\NotationName}
    \newtheorem*{notation*}{\NotationName}
    
    \theoremstyle{remark}
    \newtheorem{case}[theorem]{\CaseName}
    \newtheorem*{case*}{\CaseName}
    \newtheorem{question}[theorem]{\QuestionName}
    \newtheorem*{question*}{\QuestionName}
    \newtheorem{assumption}[theorem]{\AssumptionName}
    \newtheorem*{assumption*}{\AssumptionName}
  
  }

  \ifthenelse{\equal{\TheoremLayout}{RIGHT-HAND-NUMBERS-LAYOUT}}{%
  
    \theoremstyle{plain}
    \newtheorem{theorem}{}[\ResetCounter]
    \newtheorem*{theorem*}{\TheoremName}
    
    \newtheorem*{lemma*}{\LemmaName}
    
    \newtheorem*{corollary*}{\CorollaryName}
  
    \theoremstyle{definition}
    
    \newtheorem*{definition*}{\DefinitionName}
    
    \newtheorem*{remark*}{\RemarkName}
    
    \newtheorem*{example*}{\ExampleName}
    
    \newtheorem*{exercise*}{\ExerciseName}
    
    \newtheorem*{model*}{\ModelName}
    
    \newtheorem*{notation*}{\NotationName}
    
    \theoremstyle{remark}
    
    \newtheorem*{case*}{\CaseName}
    
    \newtheorem*{question*}{\QuestionName}
    
    \newtheorem*{assumption*}{\AssumptionName}
  
  }
  
}{}

\newcommand*{\Xop}[1]{\ensuremath{\mathrm{#1}}\xspace}

\renewcommand*{\Re}{\Xop{Re}}

\renewcommand*{\Im}{\Xop{Im}}

\DeclareMathOperator{\diag}{diag}

\DeclareMathOperator{\Ei}{Ei}

\DeclareMathAccent{\Ring}{\mathord}{operators}{"17}

\newcommand*{\Ind}[1]{\NSet{1}{#1}}

\newcommand*{\E}[1][]{\Xmath{E_{#1}}}

\newcommand*{\Cov}[1][]{\Xmath{R_{#1}}}

\newcommand*{\NSet}[3][]{\Xmath{\mathds{#2}_{#3}^{#1}}}

\newcommand*{\N}[1][]{\NSet[#1]{N}{}}

\newcommand*{\R}[1][]{\NSet[#1]{R}{}}

\newcommand*{\W}[1][]{\NSet[#1]{W}{}}

\newcommand*{\Z}[1][]{\NSet[#1]{Z}{}}

\newcommand*{\XBrace}[4][]{\Xmath{#1 #2 #4 #1 #3}}

\newcommand*{\Floor}[2][]{\XBrace[#1]{\lfloor}{\rfloor}{#2}}

\newcommand*{\RBrace}[2][]{\XBrace[#1]{(}{)}{#2}}

\newcommand*{\SBrace}[2][]{\XBrace[#1]{[}{]}{#2}}

\newcommand*{\CBrace}[2][]{\XBrace[#1]{\{}{\}}{#2}}

} %** \ifthenelse at the beginning of this file (don't remove the brackets) ****

\newtheorem{theorem}{Theorem}

\begin{document}

%\title{Ergodic Capacity of  Frequency-Selective Rayleigh Fading Channels with Correlated Scattering}
\title{\LARGE Ergodic Capacity of Discrete- and Continuous-Time, Frequency-Selective Rayleigh Fading Channels with Correlated Scattering}

\author{\authorblockN{Martin Mittelbach\authorrefmark{1}, Christian M\"uller\authorrefmark{1},
Konrad Schubert\authorrefmark{2}, and Adolf Finger\authorrefmark{1}}
\authorblockA{\authorrefmark{1}Department of Electrical Engineering and Information Technology / Communications Laboratory}
\authorblockA{\authorrefmark{2}Department of Mathematics\\
Dresden University of Technology, 01062 Dresden, Germany\\E-mail: \{mittelbach, muellerc\}@ifn.et.tu-dresden.de, schubk@math.tu-dresden.de}}

\maketitle

\def\thefootnote{}
\footnote{This research was supported by the Deutsche Forschungsgemeinschaft (DFG) under grant FI 470/7-1.}

\begin{abstract}
\label{sec:ABSTRACT}
We study the ergodic capacity of a fre\-quency-selective Rayleigh fading channel with correlated scattering, which finds application in the area of UWB. Under an average power constraint, we consider a single-user, single-antenna transmission. Coherent reception is assumed with full CSI at the receiver and no CSI at the transmitter. We distinguish between a continuous- and a discrete-time channel, modeled either as random process or random vector with generic covariance. As a practically relevant example, we  exa\-mine an exponentially attenuated Ornstein-Uhlenbeck process in detail. Finally, we give numerical results,  discuss the relation between the continuous- and the discrete-time channel model and show the significant impact of correlated scattering.
\end{abstract}

\begin{keywords}
  Frequency-selective Rayleigh fading, correlated scattering, ergodic capacity, UWB channel modeling.
\end{keywords}

\section{Introduction}
\label{sec:INTRODUCTION}

Due to the increasing role of wireless communication it is important to determine the maximum achievable information rates over multipath fading  channels. Assuming an ergodic fading process and sufficiently relaxed decoding constraints, such that fluctuations of the channel strength can be averaged out, then the \emph{ergodic capacity} is a suitable performance measure. It basically represents the average over all instantaneous channel capacities \cite{article-information-theory-en-1,book-communications-common-en-4}. 

In this paper, we examine the ergodic capacity of a single-user, single-antenna channel with full channel state information (CSI) at the receiver.  The assumption of coherent reception is reasonable if the fading is slow in the sense that the receiver is able to track the channel variations. The transmitter has no CSI but knows the statistical properties of the fading. Further, we imply an average power constraint on the channel input and Rayleigh-distributed fading. 

Under the above constraints, the ergodic capacity was investigated for the case of flat Rayleigh fading, e.g., in \cite{article-information-theory-en-22,article-information-theory-en-23,article-information-theory-en-24}. The recent interest in ultra-wideband (UWB) technologies makes it important to examine the capacity also of frequency-selective fading channels. As related information-theoretic work, consider for instance \cite{article-information-theory-en-21},\cite{article-information-theory-en-25}, where in \cite{article-information-theory-en-21} a model with two scattered components was studied. In \cite{article-information-theory-en-25} a multi-antenna system was considered, which can be adapted to serve as model for a single-antenna, frequency-selective fading channel. However, in either case uncorrelated scattering is assumed, which does \emph{not} necessarily apply to UWB channels, as repeatedly documented in the literature \cite{article-uwb-channel-en-6,article-uwb-channel-en-15,article-uwb-channel-en-17}. This is one basic difference between conventional and UWB channels. 

In this paper, we study the ergodic capacity of frequency-selective fading channels with correlated scattering. To the best of our knowledge this has not previously been  exa\-mined in the literature. We consider models appro\-priate to characterize small-scale fading effects, i.e., fading due to constructive and destructive interference of multiple signal paths. 
%, which is more relevant to the design of reliable and efficient communication systems. 
We assume the small-scale fading to be Rayleigh distributed, which is not standard in UWB channel mode\-ling \cite{article-uwb-channel-en-15}. However, detailed statistical evaluation of measurements in \cite{article-uwb-channel-en-17} \emph{support} this assumption also for UWB. Furthermore, in \cite{inproceedings-uwb-channel-en-17} it is shown that the capacity of the Rayleigh fading channel provides a tight approximation, even if other fading statistics are employed.

We distinguish between a continuous- and a discrete-time channel, where the channel impulse response (CIR) is either modeled as random process or random vector with generic covariance. Since both models are important, they are treated in parallel. The former is more general and more suitable to analysis, whereas the latter is more adequate for computer simulations and parameter estimation from measured data. Note, with the discrete approach we model equidistant samples of the CIR 
%(mathematical paths)
rather than variable-distant physical paths as in \cite{article-uwb-channel-en-8}. Modeling the sampled impulse response better describes the effective channel and is considered more robust since only aggregate physical effects need to be reflected \cite{article-uwb-channel-en-17}. This is particularly relevant where frequency-selective propagation phenomena occur. 
%effective channel (propagation channel, antennas, TX/RX filters) 

For the continuous-time Rayleigh fading channel, we exa\-mine a detailed example with special covariance. We utilize an exponentially attenuated Ornstein-Uhlenbeck process being mathematically tractable and capturing an  exponential power decay, which is common in UWB channel modeling \cite{article-uwb-channel-en-15,article-uwb-channel-en-8}.  Additionally, it incorporates exponentially correlated scattering as measured in \cite{article-uwb-channel-en-6}. 

This paper is organized as follows: Section~\ref{sec:CHANEL_MODELS} specifies the channel models, and in Section~\ref{sec:CALC_CERG} we derive respective expressions for the ergodic capacity. In Section~\ref{sec:OU_PROCESS} we analyze the example of an exponentially attenuated Ornstein-Uhlenbeck process, give numerical results, and discuss the relations between the continuous- and the discrete-time model. Section~\ref{sec:CONCLUSION} finally concludes the paper.

The following  notation is used. The operator $\E{[\cdot]}$ denotes expectation, $j$ is the imaginary unit, and $\overline{X}$ is the complex conjugate of $X$. We use the abbreviations R-integral for Riemann integral and i.i.d. for independent and identically distributed.  A wide-sense stationary random process is referred to as stationary process. We define the sets $\Z_K\text{:=}\CBrace{0,\ldots,K-1}$, $K\,$$\in\,$$\N$, and $\W\text{:=}\RBrace[\big]{\text{--}\tfrac{W}{2},\tfrac{W}{2}}$, $W > 0$. Further, $X=(X_k)$, $k\,$$\in\,$$\Z_K$, represents a column vector of size $K$ with components $X_k$. Matrix notation is equivalent using two indices. 
 
%We indicate a variable in frequency domain using the $\Hat{\quad}$-label, and $\overline{X}$ means complex conjugation of $X$. With the indices $f,n$ we distinguish between the discrete and continuous case. We say a random process is stationary if it is wide-sense stationary. We use the term R-integral to denote Riemann integral. The 
%abbreviation i.i.d. is the short hand notation for idependent and identically distributed.
%\newpage

\section{Channel models}
\label{sec:CHANEL_MODELS}

As general fading multipath channel model we consider a linear time-varying system with equivalent lowpass impulse response $(H_{\tau,t})$  being a complex random process in the time variable $t\,$$\in\,$$\R$ and the delay parameter $\tau\,$$\in\,$$\R$. Then a realization $h(\tau,t):=H_{\tau,t}(\omega)$ is the channel response at time $t$ due to an impulse at time $t-\tau$ \cite{article-information-theory-en-1,book-communications-common-en-1}. Next, for fixed $\tau$ we assume the channel to be invariant within coherence intervals of fixed length. Hence, we may  consider the random process $(H_{\tau,n})$, $\tau\,$$\in\,$$\R$, in the discrete time variable $n\,$$\in\,$$\Z$. Further, we imply $(H_{\tau,n})$ to be stationary and independent for fixed $\tau$, which corresponds to the \emph{block fading} model. As a consequence of this major simplification we are able to drop the time index $n$ and model the channel as random process $(H_\tau)$ in the delay variable $\tau\,$$\in\,$$\R$. Another widely-used assumption of uncorrelated scattering, i.e., $\E{[H_{\tau} \overline{H}_{\tau'}]}=0$ for $\tau,\tau'\,$$\in\,$$\R$ with $\tau\ne\tau'$, does not necessarily hold  \cite{article-uwb-channel-en-6,article-uwb-channel-en-15,article-uwb-channel-en-17}.  Therefore, we assume correlated scattering, which is a \emph{substantial} difference to previous work.

In addition to fading we assume additive white Gaussian noise (AWGN) at the receiver. Below, we distinguish bet\-ween a continuous- and a discrete-time channel. Thus the noise is either modeled as complex white Gaussian process $(Z_t), t\,$$\in\,$$\R$, with  i.i.d. real and imaginary part, each of zero mean and power spectral density $\tfrac{N_0}{2}$ or as complex white Gaussian process $(Z_n), n\,$$\in\,$$\Z$, with  i.i.d. real and imaginary part, each of zero mean and variance $\tfrac{N_0}{2}$.

Next, we will specify stochastic properties of $(H_{\tau})$ to obtain a Rayleigh fading channel model for continuous and discrete time.

\subsection{Continuous-Time Rayleigh Fading Channel Model}
\label{sec:CONT_MODEL}

Let $(\Tilde{X}_{\tau})$, $(\Tilde{Y}_{\tau})$, $\tau\,$$\in\,$$\R$, be real stationary i.i.d. Gaussian processes with zero mean and continuous  covariance function $\Tilde{\Cov}$. Let $g$  be an R-integrable function, i.e., $\int_{-\infty}^{\infty}g(\tau)\,d\tau$ exists as improper R-integral. We  define
\begin{equation}
\label{eqn:DECAY_FUNC}
g(\tau):=u(\tau)\Ind{[0,\infty)}(\tau) , \, \tau\in\R,
\end{equation}
with $\Ind{A}$ the indicator function  being $1$ if $\tau\,$$\in\,$$A$ and $0$ otherwise.
Then $(X_{\tau}):=(\Tilde{X}_{\tau}g(\tau))$, $(Y_{\tau}):=(\Tilde{Y}_{\tau}g(\tau))$, $\tau\,$$\in\,$$\R$, are real i.i.d. Gaussian processes with zero mean and covariance function 
\begin{equation}
\label{eqn:COV_FUNC}
\Cov(\tau,\tau') = \Tilde{\Cov}(\tau - \tau')g(\tau)g(\tau') , \,\tau,\tau' \in \R.
\end{equation}
The indicator \Ind{A} is introduced since only delays $\tau\,$$\geq\,$$0$ are meaningful and $u$ is some suitable function modeling power decay over $\tau$. Thus, $(X_{\tau})$, $(Y_{\tau})$  are attenuated, non-stationary versions of $(\Tilde{X}_{\tau})$, $(\Tilde{Y}_{\tau})$ with $X_{\tau}\,$$=\,$$Y_{\tau}\,$$=\,$$0$ for $\tau\,$$<\,$$0$. 
Finally, the continuous-time Rayleigh fading channel model is defined as $(H_\tau)\,$$:=\,$$(X_\tau + jY_\tau)$, $\tau\,$$\in\,$$\R$.%, with $j$ the imaginary unit. 

Note, by now nothing is said about band- or time-limitation of the channel. 
%Further, any real Gaussian process and its Hilbert transform are i.i.d. processes \cite{in-preparation}. This motivates the above equivalent lowpass channel model.
%In the example in Section~\ref{sec:OU_PROCESS}
Combining a stationary process with a decaying function allows us to take advantage of well-investigated random processes as described,  e.g., in \cite[Ch.~3.5/3.7]{book-mathematics-common-en-4} while capturing the decaying nature of measured CIRs. %Below we use the constant $c\,$$:=\,$$2\int_{-\infty}^{\infty}\Cov(\tau,\tau)\,d\tau$ 
%=2\Tilde{\Cov}(0)\int_{-\infty}^{\infty}g^2(\tau)\,d\tau$ 
%for normalization, 
For normalization we will use the constant $c\,$$:=\,$$2\int_{-\infty}^{\infty}\Cov(\tau,\tau)\,d\tau$, 
%=2\Tilde{\Cov}(0)\int_{-\infty}^{\infty}g^2(\tau)\,d\tau$ 
 which represents the mean energy contained in $(H_{\tau})$. It is finite if $g^2$ is R-integrable. This is obviously a reasonable assumption from a practical viewpoint. Further conditions of $\Tilde{\Cov}$ being continuous and $g$ being R-integrable we will motivate later.

Uncorrelated scattering can also be included in the continuous model utilizing a covariance function $\Tilde{\Cov}$ in form of a Dirac delta distribution, i.e.,  $\Tilde{\Cov}(\Tilde{\tau})\,$$=\,$$\Tilde{c}\delta(\Tilde{\tau})$ for some $\Tilde{c}\,$$>\,$$0$. However, to rigorously derive the ergodic capacity in this case we need to extend the  mathematical tools applied in this work, as will be briefly discussed in Section~\ref{sec:CALC_CERG}. 

%to be extended to treat this case as well. The ergodic capacity for uncorrelated scattering 
%rigorously
%we merly present res as well.

%Using the composition of a stationary process with a decaying function allows us to take advantage of the theory of stationary processes  while capturing the decaying nature of measured channel impulse responses. 
%The condition $g\in L_1(\R)$  we will motivate later.

\subsection{Discrete-Time Rayleigh Fading Channel Model}
\label{sec:DISC_MODEL}

For the discrete-time model we assume the channel to be band-limited to $\W$. Then $(H_\tau)$ can be sampled  at delays $\tau\,$$=\,$$\tfrac{l}{W},l\,$$\in\,$$\Z$, to obtain the complex random process $(H_l)$ in the discrete delay variable $l\,$$\in\,$$\Z$. Note, we have infinite expansion in delay due to band-limitation and $H_{l}\,$$=\,$$0$ for $l\,$$<\,$$0$ due to $H_{\tau}\,$$=\,$$0$ for $\tau\,$$<\,$$0$. In the following we will refer to $H_l$ as the $l$-th channel tap. Next, we approximate $(H_l)$, $l\,$$\in\,$$\Z$,  by a random vector $H\,$$:=\,$$(H_0,\ldots ,H_{L-1})$ of size \text{$L\,$$:=\,$$\Floor{ WT_d} + 1$}. Here, $L$ models the number of significant channel taps with $T_d$ being the channel delay spread. This practically feasible approximation is well-founded since CIRs are sufficiently close to zero for delays $\tau\,$$>\,$$T_d$, which is mathematically captured by $u$ in \eqref{eqn:DECAY_FUNC}. Note, for flat fading we have $L\,$$=\,$$1$, for frequency-selective fading we have $L\,$$>\,$$1$, and for UWB we clearly have $L\,$$\gg\,$$1$. Finally, we denote the $L$-dimensional complex random vector $H$ as $H\,$$=\,$$X+jY$, where $X,Y$ are real i.i.d. Gaussian vectors with zero mean and covariance matrix 
\begin{equation}
\label{eqn:COV_MAT}
\Gamma :=(\gamma_{ik}),\quad \gamma_{ik}:=\varrho_{ik} \sigma_i \sigma_k,\quad i,k \in \Z_L.
\end{equation}
%%
%Therein, $2\sigma_l^2$ is the mean power of the $l$-th channel tap $H_l$ with $\sigma_l^2:=\E\SBrace{X_l^2}=\E\SBrace{Y_l^2}=\E\SBrace{|H_l|^2}/2$. If we define  $\sigma^2:=(\sigma_0^2,\ldots,\sigma_{L-1}^2)$, then $2\sigma^2$ is the mean power delay profile,
Therein, $2\sigma_l^2$ is the mean power of the $l$-th channel tap $H_l$ with $\sigma_l^2\,$$:=\,$$\E\SBrace{X_l^2}\,$$=\,$$\E\SBrace{Y_l^2}\,$$=\,$$\E\SBrace{|H_l|^2}/2$. If we define   \text{$p_{re}\,$$:=\,$$p_{im}\,$$:=\,$$(\sigma_l^2), \,l\,$$\in\,$$\Z_L$,} then $p\,$$:=\,$$p_{re}+p_{im}$ is the mean power delay profile, which is related to $u$ in \eqref{eqn:DECAY_FUNC}. The coefficients $\varrho_{ik}$ represent the normalized correlation between tap $H_i$ and $H_k$. Uncorrelated scattering is included as a special case, where the covariance  satisfies $\Gamma\,$$=\,$$\diag(p/2)$. For norma\-lization we set the mean power of $H$ to 1, i.e., $\sum_{l=0}^{L-1}\sigma_l^2\,$$=\,$$\frac{1}{2}$. Subsequently, we refer to the above defined model as the discrete-time Rayleigh fading channel.

\section{Calculation of Ergodic Capacity}
\label{sec:CALC_CERG}

We now calculate the ergodic capacity for the defined continuous- and discrete-time Rayleigh fading channel under the general conditions specified in the beginning of the introduction. Unless stated otherwise, capacity expressions are given in [bits/s] for the continuous model and in [bits/s/Hz] for the discrete model.

\subsection{Capacity Formulae}
\label{sec:CAP_FORMULAE}

In the continuous case the ergodic capacity within the frequency band \W is calculated by \cite{article-information-theory-en-21,article-information-theory-en-1} 
\begin{equation}
\label{eqn:CERG_CONT}
C = \E\SBrace[\bigg]{\int_{-W/2}^{W/2} \log_2 \RBrace[\big]{1 + \alpha \, |\Hat{H}_{f}|^2} \, df},
\end{equation}
where $(\Hat{H}_f)\,$$:=\,$$(\int_{-\infty}^{\infty}\mathrm{e}^{-j 2 \pi f \tau}H_\tau d\tau), f$$\,\in\,$$\R$, is the Fourier transform of the process $(H_\tau)$ and $\alpha\,$$:=\,$$\tfrac{P}{N_0W}$ defines the average signal-to-noise ratio (SNR). The type of integral to calculate $C$ and $\Hat{H}_f$ is a stochastic R-integral as defined, e.g., in \cite[Ch.~3.4]{book-communications-common-en-5}. Clearly, its existence is necessary for these quantities to make sense. Expression \eqref{eqn:CERG_CONT} is valid assuming an information carrying complex envelope input signal band-limited to \W with constant power spectral density $\tfrac{P}{W}$ and mean power constraint $P$.

For the discrete-time channel model \eqref{eqn:CERG_CONT} is also applicable if the discrete-time Fourier transform (DTFT), i.e., $\Hat{H}_{f}\,$$:=\,$$\sum_{l=0}^{L-1} H_l \mathrm{e}^{-j 2 \pi fl/W}, f\,$$\in\,$$\W$, is used to calculate the spectrum of the channel vector $H$. However, with the discrete approach we aim at a numerically easy-to-compute model, preferably discrete in the frequency domain as well. Therefore, we approximate the DTFT by an $N$-point DFT, i.e., evaluating the spectrum at $N$ points. Thus we calculate $\Hat{H}_n\,$$:=\,$$\Hat{H}_{f}|_{f=nW/N}$, $n\,$$\in\,$$\Z_N$, and obtain the complex random vector $\Hat{H}\,$$:=\,$$(\Hat{H}_0,\ldots,\Hat{H}_{N-1})$. This actually means we are dividing the spectrum into $N$ flat, parallel sub-channels which corresponds to an OFDM-based system approach with $N$ sub-carriers \cite[Ch.~5.3.3/5.4.7]{book-communications-common-en-4}. The ergodic capacity is then given by
\begin{equation}
\label{eqn:CERG_DISC}
C_{N} = E\SBrace[\bigg]{\frac{1}{N} \sum_{n = 0}^{N-1}\log_2(1 + \alpha \, |\Hat{H}_n|^2)},
\end{equation}
where $\alpha$ again means average SNR, now $\alpha\,$$:=\,$$\tfrac{P}{N_0}$. Considering the parallel channels in frequency, the AWGN process at the receiver translates into a complex zero mean Gaussian random vector with independent real and imaginary parts with $N$ independent components each having variance $\tfrac{N_0}{2}$. The average power constraint of $P$ on each discrete-time channel input symbol converts to $NP$ on the set of sub-channels (per OFDM symbol). Note that $\tfrac{WC_N}{N}\rightarrow C$ as $N\rightarrow\infty$.

In the following we evaluate \eqref{eqn:CERG_CONT} and \eqref{eqn:CERG_DISC}.

\subsection{Continuous-Time Rayleigh Fading Channel}
\label{sec:CALC_CERG_CONT}

\begin{theorem}
\label{thm:CERG_CONT}
The ergodic capacity \eqref{eqn:CERG_CONT} of the continuous-time Rayleigh fading channel is given by
\begin{equation}
\label{eqn:CERG_CONT_EVAL}
C = \tfrac{1}{\ln(2)} \int_{-W/2}^{W/2} \exp\RBrace[\Big]{\tfrac{1}{2 \alpha \,\Hat{\sigma}^2(f)}} \Ei_1\RBrace[\Big]{\tfrac{1}{2 \alpha \,\Hat{\sigma}^2(f)}} \, df ,
\end{equation}
where
\begin{equation}
\label{eqn:SIGMA_HAT_CONT}
%\Hat{\sigma}^2(f) = \int_{0}^{\infty}\int_{0}^{\infty} \Cov(\tau,\tau') \cos\RBrace{2\pi (\tau-\tau') f} \, d\tau \, d\tau',\,f\in\R,
\Hat{\sigma}^2(f)\hspace{-0.4ex}=\hspace{-0.7ex}\int_{0}^{\infty}\hspace{-1.8ex}\int_{0}^{\infty}\hspace{-1ex}\Cov(\tau,\tau') \cos\RBrace{2\pi (\tau\text{\,--\,}\tau') f} \, d\tau \, d\tau',\,f\in\R,
\end{equation}
with $\Cov$ as defined in \eqref{eqn:COV_FUNC} and $\Ei_{m}(z)\,$$:=\,$$\int_{1}^{\infty}\mathrm{e}^{-tz}t^{-m}dt$. In particular, $\Ei_1(z)\,$$=\,$$\text{--}\Ei(\text{--}z)$ with $\Ei$ the exponential integral \cite[5.1]{book-mathematics-common-en-1}.
If $\Tilde{\Cov}$ in \eqref{eqn:COV_FUNC} satisfies $\Tilde{\Cov}(\Tilde{\tau})\,$$=\,$$\Tilde{c}\delta(\Tilde{\tau})$, we have uncorrelated  scattering and obtain
\begin{equation}
\label{eqn:CERG_US_CONT}
C = \tfrac{W}{\ln(2)} \exp\RBrace[\big]{\tfrac{1}{c\alpha}} \Ei_1\RBrace[\big]{\tfrac{1}{c\alpha}},
\end{equation}
where $c\,$$=\,$$2\Tilde{c}\int_{-\infty}^{\infty}g^2(\tau)\,d\tau$ is the only parameter of the CIR that has an influence on the capacity.
% only depending on the SNR $\alpha$ and the mean energy $c=2\Tilde{c}\int_{-\infty}^{\infty}g^2(\tau)\,d\tau$.
Expression \eqref{eqn:CERG_US_CONT} is well known and was already derived in \cite{article-information-theory-en-21}.

\end{theorem}

%\cite{book-communications-common-en-3,book-communications-common-en-5,book-communications-common-en-6,book-mathematics-common-en-6}

\emph{Proof:} Here we just provide the main parts of the proof that allow to understand the underlying methods. A complete proof containing omitted details is given in \cite{in-preparation}. Relevant properties of stochastic R-integrals and of mean square calculus can also be found in \cite[Ch.~2.1-2/8.1-2]{book-communications-common-en-3}, \cite[Ch.~3.3/3.4/3.6]{book-communications-common-en-5}, \cite[Ch.~8-4.]{book-communications-common-en-6}. %\cite[III.B/III.C]{book-mathematics-common-en-6}.
% an outline in terms of main propositions rigorously proved
%%

\emph{(i) Stochastic R-integral}: Let $(U_{\tau})$, $\tau\,$$\in\,$$\R$, be a complex random process with $\E{[|U_{\tau}|^2]}\,$$<\,$$\infty$. Then it can be shown that the stochastic R-integral $I_U\,$$:=\,$$\int_{-\infty}^{\infty}U_{\tau}\,d\tau$ exists if and only if $\E{[U_{\tau}\overline{U}_{\tau'}]}$ is R-integrable, i.e., 
\begin{equation}
\label{eqn:COND_RINTBLE}
\int_{-\infty}^{\infty}\int_{-\infty}^{\infty}\E{[U_{\tau}\overline{U}_{\tau'}]} \, d\tau \, d\tau'
\end{equation}
exists as improper R-integral. Then we obtain
\begin{equation}
\label{eqn:E_RINT}
\E{[I_U]}=\int_{-\infty}^{\infty}\E{[U_{\tau}]} \, d\tau.
\end{equation}
If we have another complex process $(V_{\tau})$, $\tau\,$$\in\,$$\R$, with $\E{[|V_{\tau}|^2]}\,$$<\,$$\infty$ for which $I_V\,$$:=\,$$\int_{-\infty}^{\infty}V_{\tau}\,d\tau$ exists, then it can further be shown that 
\begin{equation}
\label{eqn:COV_RINT}
\E{[I_U\overline{I}_V]}=\int_{-\infty}^{\infty}\int_{-\infty}^{\infty}\E{[U_{\tau}\overline{V}_{\tau'}]} \, d\tau \, d\tau'.
\end{equation}

\emph{(ii) Existence of $(\Hat{H}_f)$}: We use \eqref{eqn:COND_RINTBLE} to prove that the Fourier transform $(\Hat{H}_f)\,$$=\,$$(\int_{-\infty}^{\infty}\mathrm{e}^{-j 2 \pi f \tau}H_\tau d\tau)$, $f\,$$\in\,$$\R$, of the CIR $({H}_\tau)$, $\tau\,$$\in\,$$\R$, exists. Thus we set $U_{\tau}\,$$:=\,$$\mathrm{e}^{-j 2 \pi f \tau}H_\tau$ and obtain \eqref{eqn:COND_RINTBLE} with $\E{[U_{\tau}\overline{U}_{\tau'}]}\,$$=\,$$2\Cov(\tau,\tau')\mathrm{e}^{-j 2 \pi f \tau}\mathrm{e}^{j 2 \pi f \tau'}$, where $\Cov$ is the covariance given in \eqref{eqn:COV_FUNC}. If $\Cov$ in \eqref{eqn:COV_FUNC} is such that $\Tilde{\Cov}$ is  continuous and $g$ is R-integrable then  \eqref{eqn:COND_RINTBLE} exists for all $f\,$$\in\,$$\R$ and hence $(\Hat{H}_f)$. Note, for $(\Hat{H}_f)$ to exist, we can alternatively require $\Cov$ to be R-integrable.

\emph{(iii) Distribution of $(\Hat{H}_f)$}: It can be shown that any linear transformation of a Gaussian process is a Gaussian process. Thus $(\Hat{H}_f)$ is a complex Gaussian process composed of two real Gaussian processes $(\Hat{X}_f)\,$$:=\,$$(\Re{[\Hat{H}_f]})$ and $(\Hat{Y}_f)\,$$:=\,$$(\Im{[\Hat{H}_f]})$. The process $(\Hat{X}_f)$ has zero mean, which follows from \eqref{eqn:E_RINT} with $U_{\tau}\,$$:=\,$$\Re{[\mathrm{e}^{-j 2 \pi f \tau}H_\tau]}$ and from $\E{[H_{\tau}]}\,$$=\,$$0$. We  calculate the covariance of $(\Hat{X}_f)$ using \eqref{eqn:COV_RINT} with 
%$U_{\tau}:=\Re{[\mathrm{e}^{-j 2 \pi f \tau}H_\tau]}$ 
$U_{\tau}$ as before
and $V_{\tau}\,$$:=\,$$\Re{[\mathrm{e}^{-j 2 \pi f' \tau}H_\tau]}$ resulting in $\E{[\Hat{X}_{f}{\Hat{X}}_{f'}]}\,$$=\,$$\int_{0}^{\infty}\int_{0}^{\infty} \Cov(\tau,\tau') \cos\RBrace{2\pi (\tau f-\tau' f')} \, d\tau\,d\tau'$ for $f,f'\,$$\in\,$$\R$ and $\Cov$ as in \eqref{eqn:COV_FUNC}. Correspondingly, we show with $\Re{[\cdot]}$ replaced by $\Im{[\cdot]}$ that the process $(\Hat{Y}_f)$ has zero mean and identical covariance. Again we use \eqref{eqn:COV_RINT} with $U_{\tau}\,$$:=\,$$\Re{[\mathrm{e}^{-j 2 \pi f \tau}H_\tau]}$ and $V_{\tau}\,$$:=\,$$\Im{[\mathrm{e}^{-j 2 \pi f' \tau}H_\tau]}$ to determine the cross-covariance bet\-ween $(\Hat{X}_f)$ and $(\Hat{Y}_f)$. We obtain 
%The cross-covariance bet\-ween $(\Hat{X}_f)$ and $(\Hat{Y}_f)$ is given by
% $\E{[\Hat{X}_{f}{\Hat{Y}}_{f'}]}=-\E{[\Hat{X}_{f'}{\Hat{Y}}_{f}]}$ with  $\E{[\Hat{X}_{f}{\Hat{Y}}_{f'}]}=\int_{0}^{\infty}\int_{0}^{\infty} \Cov(\tau,\tau') \sin\RBrace{2\pi (\tau f-\tau' f')} \, d\tau  d\tau'$ for $f,f'\in\R$.  Thus 
% $(\Hat{X}_f)$ and $(\Hat{Y}_f)$ are identically distributed processes but they are not independent.
$\E{[\Hat{X}_{f}{\Hat{Y}}_{f'}]}\,$$=\,$$\int_{0}^{\infty}\int_{0}^{\infty} \Cov(\tau,\tau') \sin\RBrace{2\pi (\tau f-\tau' f')} \, d\tau\,d\tau'$ for $f,f'\,$$\in\,$$\R$ and $\E{[\Hat{X}_{f}{\Hat{Y}}_{f'}]}\,$$=\,$$-$$\E{[\Hat{X}_{f'}{\Hat{Y}}_{f}]}$.  Thus the processes
 $(\Hat{X}_f)$, $(\Hat{Y}_f)$ are identically distributed but not independent. 
 
\emph{(iv) Distribution of $|\Hat{H}_f|^2$}:
From  (iii) it follows for any  $f\,$$\in\,$$\R$ that $\Hat{X}_f,\Hat{Y}_f$ are i.i.d. Gaussian random variables having zero mean and variance $\E{[\Hat{X}_{f}^2]}\,$$=\,$$\Hat{\sigma}^2(f)$ with $\Hat{\sigma}^2(f)$ as in \eqref{eqn:SIGMA_HAT_CONT}. Thus $|\Hat{H}_f|^2\,$$=\,$$\Hat{X}_f^2 + \Hat{Y}_f^2$ is an exponentially distributed random variable  with probability density function $q_f(\Hat{z})\,$$=\,$$(2\Hat{\sigma}^2(f))^{-1}\exp[(2\Hat{\sigma}^2(f))^{-1}\Hat{z}]$, $\Hat{z}\geq 0$.

%\marginpar{\scriptsize may add non-zero mean proc.?!}

\emph{(v) Existence and calculation of $C$}: The criterion \eqref{eqn:COND_RINTBLE} for a stochastic R-integral to exist and the properties \eqref{eqn:E_RINT}, \eqref{eqn:COV_RINT} are also valid for finite integration boundaries. Thus if we set $U_f\,$$:=\,$$\log_2 (1 + \alpha \, |\Hat{H}_{f}|^2)$ then $I_U$$\,:=\,$$\int_{-W/2}^{W/2}U_f \,df$ exists if and only if $\E{[{U}_{f}{{U}}_{f'}]}$ is R-integrable over $\W^2$. Due to the finite integration boundaries it is sufficient to show that $\E{[{U}_{f}{{U}}_{f'}]}$ is continuous. This is equivalent to $\E{[{U}_{f}^2]}$ being continuous. The continuity of $\E{[{U}_{f}^2]}$ can either be shown directly or equivalently by showing that the process $(U_f)$ is continuous in mean square. Now we can apply \eqref{eqn:E_RINT} and obtain $C$$\,=\,$$\E{[I_U]}\,$$=\,$$\int_{-W/2}^{W/2}\E{[U_f]} \,df$. We  calculate $\E{[U_f]}$ by evaluating the integral $\E{[U_f]}\,$$=\,$$\int_{0}^{\infty}\log_2 (1 + \alpha \, \Hat{z})q_f(\Hat{z})\,d\Hat{z}$ with $q_{f}$ as in (iv). Finally,  this yields \eqref{eqn:CERG_CONT_EVAL}.

\emph{(vi) Uncorrelated scattering}: The covariance $\Cov$ in \eqref{eqn:COV_FUNC} becomes $\Cov(\tau,\tau')$$\,=\,$$\Tilde{c}\delta(\tau - \tau')g(\tau)g(\tau')$ for $\Tilde{\Cov}(\Tilde{\tau})$$\,=\,$$\Tilde{c}\delta(\Tilde{\tau})$. Plugging $\Cov$ into \eqref{eqn:SIGMA_HAT_CONT} and using the fundamental property of the Dirac distribution of $\int_{-\infty}^{\infty}h(\tau)\delta(\tau - \tau')\,d\tau$$\,=\,$$h(\tau')$ yields $\Hat{\sigma}^2(f)$$\,=\,$$\Tilde{c}\int_{-\infty}^{\infty}g^2(\tau)\,d\tau$, independent of $f$. Then \eqref{eqn:CERG_CONT_EVAL} immediately implies \eqref{eqn:CERG_US_CONT}. Note, this is only a formal derivation.  The introduced mathematical tools are not applicable to random processes with Dirac-type covariance functions, which do not even exist in the classical sense. A rigorous mathematical treatment involves stochastic differential equations. \hfill {\footnotesize $\blacksquare$}

\subsection{Discrete-Time Rayleigh Fading Channel}

\begin{theorem}
\label{thm:CERG_DISC}
The ergodic capacity \eqref{eqn:CERG_DISC} of the discrete-time Rayleigh fading channel is given by
\begin{equation}
\label{eqn:CERG_DISC_EVAL}
C_N = \tfrac{1}{N\ln(2)} \sum_{n=0}^{N-1} \exp\RBrace[\Big]{\tfrac{1}{2 \alpha \,\Hat{\sigma}_{n}^2}} \Ei_1\RBrace[\Big]{\tfrac{1}{2 \alpha \,\Hat{\sigma}_{n}^2}}
\end{equation}
where 
\begin{equation}
\label{eqn:SIGMA_HAT_DISC}
\Hat{\sigma}_{n}^2 = \sum_{i=0}^{L-1}\sum_{k=0}^{L-1} \gamma_{ik} \cos\RBrace{2\pi (i-k) n/N}, \, n\in\Z_N,
\end{equation}
with $\gamma_{ik}$ as in \eqref{eqn:COV_MAT}. If \eqref{eqn:COV_MAT} satisfies $\Gamma\text{=}\diag(\sigma_0^2,\ldots,\sigma_{L-1}^2)$, we have uncorrelated  channel taps (uncorrelated scattering) and obtain
\begin{equation}
\label{eqn:CERG_US_EVAL}
C_N = \tfrac{1}{\ln(2)} \exp\RBrace[\big]{\tfrac{1}{\alpha}} \Ei_1\RBrace[\big]{\tfrac{1}{\alpha}}=:C_{\mathrm{us}},
\end{equation}
independent of $L, N$, and the mean power delay profile $(\sigma_0^2,\ldots,\sigma_{L-1}^2)$. Expression \eqref{eqn:CERG_US_EVAL} is identical to the capacity of the flat Rayleigh fading case as derived in \cite{article-information-theory-en-22,article-information-theory-en-23}.
\end{theorem}

%%\emph{Proof:} Theorem~\ref{thm:CERG_DISC} is similarly proved as Theorem~\ref{thm:CERG_CONT} but with less effort.
%%The existence of $\Hat{H}:=(\Hat{H}_0,\ldots,\Hat{H}_{N-1})$ is evident and exchanging summation and expectation follows directly from the linearity of \E[]. For the distribution of $|\Hat{H}_n|^2$ we have: If $\Hat{X}:=\Re{[\Hat{H}]}$, $\Hat{Y}:=\Im{[\Hat{H}]}$ (componentwise), then  $\Hat{X},\Hat{Y}$ are real i.i.d. Gaussian vectors with zero mean and covariance matrix $\Hat{\Gamma} :=(\Hat{\gamma}_{nn'})$, ${{n,n' \in \Z_N}}$, where $\Hat{\gamma}_{nn'}= \sum_{i=0}^{L-1}\sum_{k=0}^{L-1} \gamma_{ik} \cos\RBrace{2\pi (in-kn')/N }$ with $\gamma_{ik}$ as in \eqref{eqn:COV_MAT}. In particular, $\Hat{X}_n,\Hat{Y}_n$ are i.i.d. Gaussian random variables with zero mean and variance $\Hat{\sigma}^2_n:=\Hat{\gamma}_{nn}$ for $n \in \Z_N$. Thus, $|\Hat{H}_n|^2 = \Hat{X}_n^2 + \Hat{Y}_n^2$ is an exponentially distributed random variable with parameter $\frac{1}{2\Hat{\sigma}_n^2}$. The rest of part (iii) applies accordingly. 
%%
%%Finally, if we have uncorrelated scattering, i.e., $\Gamma$ is diagonal, then $\Hat{\sigma}^2_n=\sum_{l=0}^{L-1}\sigma_l^2 = \frac{1}{2}$ for $n \in \Z_N$, and we get \eqref{eqn:CERG_US_EVAL} from \eqref{eqn:CERG_DISC_EVAL}. \hfill {\footnotesize $\blacksquare$}

\emph{Proof:} Theorem~\ref{thm:CERG_DISC} is similarly proved as Theorem~\ref{thm:CERG_CONT} but with less effort. We may write the $N$-point DFT $\Hat{H}$ of the channel vector $H$ using matrix notation. Let $\Phi$$\,:=\,$$(\varphi_{nl})$ with $\varphi_{nl}$$\,:=\,$$\mathrm{e}^{-j 2 n l / N}$, $n$$\,\in\,$$\Z_N$, $l$$\,\in\,$$\Z_L$, then $\Hat{H}$$\,=\,$$\Phi H$. The existence of $\Hat{H}$ is now evident. Further, transforming a complex Gaussian vector this way yields a complex Gaussian vector \cite[7.5-2]{book-communications-common-en-3} with $\Hat{X}$$\,:=\,$$\Re{[\Hat{H}]}$, $\Hat{Y}$$\,:=\,$$\Im{[\Hat{H}]}$ being real Gaussian vectors.  We calculate means and covariances using the linearity of $\E{[\cdot]}$ and the independence of $X$$\,=\,$$\Re{[{H}]}$ and $Y$$\,=\,$$\Im{[{H}]}$. We obtain that $\Hat{X}$, $\Hat{Y}$ are identically distributed with zero mean and covariance $\E{[\Hat{X}_n\Hat{X}_{n'}]}$$\,=\,$$ \sum_{i=0}^{L-1}\sum_{k=0}^{L-1} \gamma_{ik} \cos\RBrace{2\pi (in-kn')/N }$ for $n,n'$$\,\in\,$$\Z_N$ and $\gamma_{ik}$ as in \eqref{eqn:COV_MAT}. The cross-covariance is given by $\E{[\Hat{X}_n\Hat{Y}_{n'}]}$$\,=\,$$\sum_{i=0}^{L-1}\sum_{k=0}^{L-1} \gamma_{ik} \sin\RBrace{2\pi (in-kn')/N }$ for $n,n'$$\,\in\,$$\Z_N$ with $\E{[\Hat{X}_n\Hat{Y}_{n'}]}$$\,=\,$$-$$\E{[\Hat{X}_{n'}\Hat{Y}_{n}]}$. It follows for all $n$$\,\in\,$$\Z_N$ that $\Hat{X}_n$, $\Hat{Y}_n$ are i.i.d. Gaussian random variables with zero mean and variance $\E{[\Hat{X}_{n}^2]}$$\,=\,$$\Hat{\sigma}^2_n$ with $\Hat{\sigma}^2_n$ as in \eqref{eqn:SIGMA_HAT_DISC}. Therefore, {$|\Hat{H}_n|^2$$\,=\,$$\Hat{X}_n^2 + \Hat{Y}_n^2$} is again an exponentially distributed random variable with parameter $(2\Hat{\sigma}_n^2)^{-1}$. Now we exchange summation and expectation in \eqref{eqn:CERG_DISC} and evaluate $\E{[\log_2 (1 + \alpha \, |\Hat{H}_{n}|^2)]}$ to obtain \eqref{eqn:CERG_DISC_EVAL}. 
Finally, if we have uncorrelated scattering, i.e., $\Gamma$ in \eqref{eqn:COV_MAT} is diagonal, then $\Hat{\sigma}^2_n$$\,=\,$$\sum_{l=0}^{L-1}\sigma_l^2$$\,=\,$$\frac{1}{2}$ for $n \in \Z_N$, and  \eqref{eqn:CERG_DISC_EVAL} implies \eqref{eqn:CERG_US_EVAL}. \hfill {\footnotesize $\blacksquare$}
%any linear 
% to as can be represented by matrix multiplication. with as defined in Section~\erref{sec:CAP_FORMULAE}.
%\E{[\Hat{Y}_n\Hat{Y}_{n'}]} =
%

%The existence of $\Hat{H}:=(\Hat{H}_0,\ldots,\Hat{H}_{N-1})$ is evident and exchanging summation and expectation follows directly from the linearity of \E{[]}. For the distribution of $|\Hat{H}_n|^2$ we have: If $\Hat{X}:=\Re{[\Hat{H}]}$, $\Hat{Y}:=\Im{[\Hat{H}]}$ (componentwise), then  $\Hat{X},\Hat{Y}$ are real i.i.d. Gaussian vectors with zero mean and covariance matrix $\Hat{\Gamma} :=(\Hat{\gamma}_{nn'})$, ${{n,n' \in \Z_N}}$, where $\Hat{\gamma}_{nn'}= \sum_{i=0}^{L-1}\sum_{k=0}^{L-1} \gamma_{ik} \cos\RBrace{2\pi (in-kn')/N }$ with $\gamma_{ik}$ as in \eqref{eqn:COV_MAT}. 
%
%In particular, $\Hat{X}_n,\Hat{Y}_n$ are i.i.d. Gaussian random variables with zero mean and variance $\Hat{\sigma}^2_n:=\Hat{\gamma}_{nn}$ for $n \in \Z_N$. Thus, $|\Hat{H}_n|^2 = \Hat{X}_n^2 + \Hat{Y}_n^2$ is an exponentially distributed random variable with parameter $\frac{1}{2\Hat{\sigma}_n^2}$. The rest of part (iii) applies accordingly. 

\section{Example: Exponentially attenuated Ornstein-Uhlenbeck process }
\label{sec:OU_PROCESS}

Now, we consider an example for the continuous-time Rayleigh fading channel with special covariance. We use an exponentially attenuated Ornstein-Uhlenbeck process capturing an exponential power decay. This is a common assumption in UWB channel modeling and was justified by measurements \cite{article-uwb-channel-en-15,article-uwb-channel-en-8}. In addition, it incorporates exponentially correlated scattering. Based on UWB channel measurements, a similar correlation model was used in \cite{article-uwb-channel-en-6}. Finally, we discuss relations to the discrete-time model.

\subsection{Definition}

A stationary Ornstein-Uhlenbeck process is a real Gaussian processes with zero mean and covariance function $\Tilde{\Cov}(\tau - \tau')$$\,:=\,$$\tfrac{d}{2 a} \mathrm{e}^{- a |\tau-\tau'|}$ for $\tau,\tau'$$\,\in\,$$\R$ with parameters $a,d$$\,>\,$$0$ \cite[Ch. 3.7.2/3.7.3]{book-mathematics-common-en-4}. Using the notation of Section~\ref{sec:CONT_MODEL} we set $(\Tilde{X}_\tau)$, $(\Tilde{Y}_\tau)$ to be normalized Ornstein-Uhlenbeck processes with $d$$\,:=\,$$2a$ and specify the attenuation function $u$ in \eqref{eqn:DECAY_FUNC} as $u(\tau)$$\,:=\,$$\sqrt{bc}\,\mathrm{e}^{-b\tau}$ for $\tau$$\,\in\,$$\R$ with parameters $b,c$$\,>\,$$0$, where $c$ is the constant defined in Section~\ref{sec:CONT_MODEL}. We obtain the attenuated Ornstein-Uhlenbeck processes $(X_\tau)$$\,=\,$$(\Tilde{X}_\tau g(\tau))$, $(Y_\tau)$$\,=\,$$(\Tilde{Y}_\tau g(\tau))$, $\tau$$\,\in\,$$\R$,  representing independent real and imaginary part of $(H_{\tau})$, each with covariance function 
\begin{equation}
\label{eqn:COV_ATT_OU} 
 \Cov(\tau,\tau') = 
 c \mathrm{e}^{- a |\tau-\tau'|} b \mathrm{e}^{-b(\tau+\tau')}\Ind{[0,\infty)}(\tau)\Ind{[0,\infty)}(\tau'),
\end{equation}
for $\tau,\tau'$$\,\in\,$$\R$.

\subsection{Analytical Calculations}

The expressions given in this subsection are derived in the Appendix in condensed form. %For details and alternative representations of \eqref{eqn:CERG_OU_EVAL} please refer to \cite{in-preparation}.

Calculating \eqref{eqn:SIGMA_HAT_CONT} using \eqref{eqn:COV_ATT_OU} we get 
\begin{equation}
\label{eqn:SIGMA_HAT_ATT_OU}
\Hat{\sigma}^2(f) = \frac{c(a+b)}{(a+b)^2+(2\pi f)^2}
\end{equation}
by elementary integration. One representation of the closed form solution of \eqref{eqn:CERG_CONT_EVAL} is given by
\begin{equation}
\label{eqn:CERG_OU_EVAL}
C = \tfrac{W}{\ln(2)} \sum_{n=0}^{\infty} \tfrac{1}{n+1}\sum_{k=0}^{n}\tfrac{1}{k + 2} L_k^{\frac{1}{2}}\RBrace{\beta_1W^2/4}\,L_{n-k}^{-\frac{1}{2}}\RBrace{\beta_1\beta_2},
\end{equation}
where $L_k^\mu$ is the generalized Laguerre polynomial of order $k$ \cite[8.970]{book-mathematics-common-en-2},  $\beta_1$$\,:=\,$$2\pi^2/(\alpha c(a+b))$, and $\beta_2$$\,:=\,$$(a+b)^2/(4\pi^2)$. 

Upper and lower bounds for \eqref{eqn:CERG_OU_EVAL} are given by
\begin{align}
\label{eqn:CERG_OU_BOUNDS} 
\nonumber
C_\theta =\,& W\log_2\RBrace[\big]{1 + 2\mathrm{e}^{-\theta}\alpha\Hat{\sigma}^2(W/2)} - \tfrac{4\sqrt{\beta_2}}{\ln(2)}\arctan\RBrace[\Big]{\tfrac{W/2}{\sqrt{\beta_2}}}\\
          &+ \tfrac{4\sqrt{\beta_2+\mathrm{e}^{-\theta}/\beta_1}}{\ln(2)}\arctan\RBrace[\bigg]{\tfrac{W/2}{\sqrt{\beta_2+\mathrm{e}^{-\theta}/\beta_1}}},
\end{align}
where we have an upper bound for $\theta$$\,=\,$$0$ and a lower bound for $\theta$$\,=\,$$\gamma$ with $\gamma$ being the Euler constant. 

In case the effect of the channel  outside the band \W is negligible, i.e., $W$ is sufficiently large (depending on the parameters $a,b,c,\alpha$), then \eqref{eqn:CERG_OU_EVAL} can be closely  approximated by integrating \eqref{eqn:CERG_CONT_EVAL}  with \eqref{eqn:SIGMA_HAT_ATT_OU} over entire $\R$ yielding
\begin{equation}
\label{eqn:CERG_OU_APPROX}
C_\approx = \tfrac{\pi}{\ln(2)\sqrt{\beta_1}} \exp(\beta_1\beta_2) \Gamma (\tfrac{1}{2},\beta_1\beta_2),
\end{equation}
where $\Gamma (\mu,z)$$\,:=\,$$\int_{z}^{\infty}\mathrm{e}^{-t}t^{\mu-1}dt$ is the incomplete gamma function \cite[8.350.2]{book-mathematics-common-en-2}.

Note, $C_\approx$ in \eqref{eqn:CERG_OU_APPROX} is just an approximate expression for proper parameter ranges but not the capacity for infinite bandwidth, i.e., not $\lim_{W\rightarrow\infty}C$ since $\alpha$ and thus $\beta_1$ depend on $W$ as well.

\subsection{Numerical Results and Relation to the Discrete-Time Rayleigh Fading Channel}
\label{sec:NUM_RESULTS}

Here, we give numerical examples for the previously derived expressions and show the connection to the discrete-time channel. To do so, we first define two constants for the continuous-time channel. Let $\Hat{\varepsilon}$$\,\in\,$$(0,1)$ be the portion of the channel within the frequency band \W in terms of mean energy, i.e., $\Hat{\varepsilon}$$\,:=\,$$\tfrac{2}{c}\int_{-W/2}^{W/2}\Hat{\sigma}^2(f)\,df$, where $c$ is defined in Section~\ref{sec:CONT_MODEL}. If we fix $\Hat{\varepsilon}$, then we get $W$$\,=\,$$\tfrac{a+b}{\pi}\tan(\tfrac{\pi}{2}\Hat{\varepsilon})$ by elementary integration. Further let $\varepsilon$$\,\in\,$$(0,1)$ be the portion of the channel within the time interval $[0,T_d]$ in terms of mean energy, i.e., $\varepsilon$$\,:=\,$$\tfrac{2}{c}\int_{0}^{T_d}\Cov(s,s)\,ds$. If we fix $\varepsilon$, then we get $T_d$$\,=\,$$\tfrac{-1}{2b}\ln(1-\varepsilon)$ by elementary integration.

Consider the continuous-time channel within the frequency band \W. To represent this channel by the discrete-time model we sample the covariance function $\Cov$ of \eqref{eqn:COV_ATT_OU} over the range $[0,T_d]$$\,\times\,$$[0,T_d]$ by $\frac{1}{W}$-spacing to get the covariance matrix $\Gamma$ of \eqref{eqn:COV_MAT}. We normalize $\Gamma$ to have channel vectors with unit mean energy as stated in Section~\ref{sec:DISC_MODEL}. In addition, we choose $\varepsilon$ to be close to 1, i.e., truncation in time is negligible. Note, this discrete version of the channel resembles the behavior of the continuous channel in time domain. However, the spectrum is different due to aliasing since the Ornstein-Uhlenbeck process is not band-limited because of \eqref{eqn:SIGMA_HAT_ATT_OU}. This, in turn, is negligible if $\Hat{\varepsilon}$ is close to 1. To obtain equality in frequency domain, we have to sample a lowpass-filtered version of the continuous channel, which of course has different behavior in time. Since we aim at equivalence in time, we consider the former approach. Finally, for the discrete and the continuous model to be comparable we have to set $c$$\,:=\,$$W/\Hat{\varepsilon}$.

\begin{figure}[htbp]
  \centering
  \vspace{-2ex}
  \includegraphics[width=8.0cm]{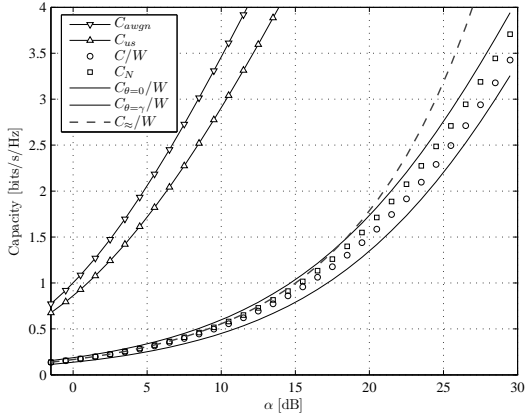}  
  \vspace{-2ex}
  \caption{Ergodic capacity Example~1: $a=b=\frac{1}{2}$, $\varepsilon=0.998$, $\Hat{\varepsilon}=0.998$.}
  \label{fig:Cap1}
\end{figure}

\begin{figure}[htbp]
  \centering
  \includegraphics[width=8.0cm]{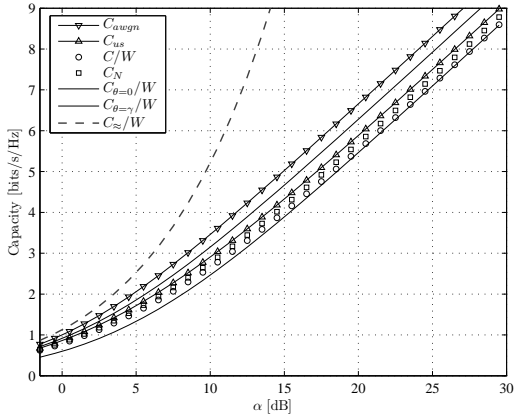}  
  \vspace{-2ex}
  \caption{Ergodic capacity Example~2: $a=b=\frac{1}{2}$, $\varepsilon=0.998$, $\Hat{\varepsilon}=0.800$.}
  \label{fig:Cap2}
\end{figure}

As a numerical example, we set $a$$\,:=\,$$b$$\,:=\,$$\tfrac{1}{2}$, and $\varepsilon$$\,:=\,$$0.998$  resulting in $T_d$$\,=\,$$6.2146$, when normalized to seconds. In Example~1 (Fig.~\ref{fig:Cap1}), we set $\Hat{\varepsilon}$$\,:=\,$$0.998$ leading to $W$$\,=\,$$101.32$, when normalized to $\text{[Hz]}$, $c$$\,=\,$$101.52$, and $L$$\,=\,$$630$. In Example~2 (Fig.~\ref{fig:Cap2}), we set $\Hat{\varepsilon}$$\,:=\,$$0.800$ leading to $W$$\,=\,$$0.9796$, $c$$\,=\,$$1.2245$, and $L$$\,=\,$$7$. Capacity expressions are considered as function of $\alpha$ and are given in [bits/s/Hz], where normalization to the respective bandwidth $W$ is performed if required. The values of $C/W$ are obtained by numerically evaluating \eqref{eqn:CERG_CONT_EVAL} and $C_N$ is computed with $N$$\,=\,$$6300$. As reference curves the AWGN channel capacity $C_{\mathrm{awgn}}$$\,=\,$$\log_2(1+\alpha)$ is plotted. Another reference is the capacity $C_{\mathrm{us}}$ for uncorrelated channel taps \eqref{eqn:CERG_US_EVAL}, which is an upper bound for the correlated case due to Jensen's inequality.

We observe differences between $C_N$ and $C/W$ which are due to the applied approximations, i.e., truncation in time, aliasing, and discretization of spectrum. They are prima\-rily minor, particularly in Example~2, but increase with $\alpha$. The distance to $C_{\mathrm{us}}$ is considerable in Fig.~\ref{fig:Cap1} but small in Fig.~\ref{fig:Cap2}. This depends on the concentration of the spectrum within the considered band, which in turn is controlled by the degree of correlation (parameter $a$) and the delay spread (parameter $b$). The tightness of the bounds is parameter-dependent. Especially the lower bound in Fig.~\ref{fig:Cap2} is very tight for $\alpha$$\,>\,$$15\text{dB}$. Finally, $C$ is well approximated by $C_{\approx}$ in Example~1 for  $\alpha$$\,<\,$$15\text{dB}$ but is inappropriate in Example~2, since $\Hat{\varepsilon}$ is not close enough to 1.
%%
%%%
%\begin{figure}[htbp]
%  \centering
%  \epsfig{file=./figures/eps/cap1.eps,width=8.2cm}  
%  \caption{Ergodic capacity Example~1: $a=b=\frac{1}{2}$, $\Hat{\varepsilon}=0.998$, $\varepsilon=0.998$}
%  \label{fig:Cap1}
%\end{figure}
%
%\begin{figure}[htbp]
%  \centering
%  \epsfig{file=./figures/eps/cap2.eps,width=8.2cm}  
%  \caption{Ergodic capacity Example~2: $a=b=\frac{1}{2}$, $\Hat{\varepsilon}=0.998$, $\varepsilon=0.800$}
%  \label{fig:Cap2}
%\end{figure}
%%%%
%%

\section{Conclusion}
\label{sec:CONCLUSION}

In this paper, we determined the ergodic capacity of a frequency-selective Rayleigh fading channel with correlated scattering. We considered a  continuous- and a discrete-time channel and examined a detailed example incorporating exponential power decay and exponentially correlated scattering. Analytical, approximate, and bounding expressions as well as numerical results were presented and the relation between continuous- and discrete-time models were discussed. The results illustrate significant differences between the capacities for correlated and uncorrelated scattering.
Future work includes for example: considering other stationary processes, assuming non-perfect CSI, further elaborating \eqref{eqn:CERG_OU_EVAL}, analyzing the outage capacity, or estimating model parameters from measured data.

\begin{appendix}
\label{sec:APPENDIX}

\subsection{Derivation of \eqref{eqn:CERG_OU_EVAL}:}
\begin{enumerate}
	\item Using \eqref{eqn:SIGMA_HAT_ATT_OU} in \eqref{eqn:CERG_CONT_EVAL} and $\Ei_1(x)\mathrm{e}^x=\sum_{n=0}^{\infty}\frac{1}{n+1}L_n(x)$ \cite[5.11.1.4]{book-mathematics-common-en-3} we get $C\text{=}\tfrac{1}{\ln(2)}\sum_{n=0}^{\infty}\tfrac{1}{n+1}I_1$ with $I_1:=\int_{-W/2}^{W/2}L_n(\beta_1(\beta_2+ f^2))\,df$ by Lebesgue's theorem.
	\item Using the substitution $s=\beta_1f^2$ and $L_n(x+y)=\sum_{k=0}^n L_k^{-\frac{1}{2}}(x)L_{n-k}^{-\frac{1}{2}}(y)$ \cite[4.4.2.3]{book-mathematics-common-en-3} then yields $I_1 = \beta_1^{-\frac{1}{2}}\sum_{k=0}^{n}L_{n-k}^{-\frac{1}{2}}(\beta_1\beta_2)I_2$ with the remaining integral $I_2:=\int_{0}^{\beta_1(W/2)^2}s^{-\frac{1}{2}}L_k^{-\frac{1}{2}}(s)\,ds$.
	\item Finally, we evaluate the term $I_2$ using $\int_0^t s^\mu L_k^\mu(s)\, ds = \tfrac{1}{k+\mu+1}t^{\mu+1}L_k^{\mu+1}(t)$ \cite[1.14.3.4]{book-mathematics-common-en-3} to get \eqref{eqn:CERG_OU_EVAL}.	
\end{enumerate}

%Remark: Using \cite[]{} and \cite[]{} we 

\subsection{Derivation of \eqref{eqn:CERG_OU_BOUNDS}:}
We simply use the inequalities $\ln(1+\tfrac{\mathrm{e}^{-\gamma}}{x}) < \Ei_1(x)\mathrm{e}^x$  \cite[eq. (13)]{inproceedings-information-theory-en-6} and  $\Ei_1(x)\mathrm{e}^x<\ln(1+\tfrac{1}{x})$ \cite[5.1.20]{book-mathematics-common-en-1} to replace the integrand. Then elementary integration and  the monotony of the integral yields \eqref{eqn:CERG_OU_BOUNDS}.

\subsection{Derivation of \eqref{eqn:CERG_OU_APPROX}:}
\begin{enumerate}
	\item Using \eqref{eqn:SIGMA_HAT_ATT_OU} and infinite integration boundaries in \eqref{eqn:CERG_CONT_EVAL}  we get $C_\approx = \frac{\mathrm{e}^{\beta_1\beta_2}}{\ln(2)\sqrt{\beta_1}}\int_0^\infty \frac{\mathrm{e}^{s}}{\sqrt{s}}\Ei_1(s+\beta_1\beta_2)\,ds$ by again substituting $s=\beta_1f^2$.
	
	\item With $\int_0^\infty \frac{\mathrm{e}^t}{t^{1-\mu}}\Ei_1(t+\nu)\,dt = \frac{\pi\nu^{(\mu-1)/2}\mathrm{e}^{-\nu/2}}{\sin(\mu\pi)}W_{\frac{\mu-1}{2},\frac{\mu}{2}}(\nu)$ \cite[2.5.3.14]{book-mathematics-common-en-3} and $W_{-\frac{1}{4},\frac{1}{4}}(y)=y^{\frac{1}{4}}\mathrm{e}^{\frac{y}{2}}\Gamma(\frac{1}{2},y)$ \cite[9.236.1, 8.359.3]{book-mathematics-common-en-2},
	where $W_{\kappa,\lambda}$ is the Whittaker's W-function \cite[9.220]{book-mathematics-common-en-2}, we finally get \eqref{eqn:CERG_OU_APPROX}.
\end{enumerate}

\end{appendix}

\section*{Acknowledgment}
% optional entry into table of contents (if used)
%\addcontentsline{toc}{section}{Acknowledgment}
The authors wish to thank Lothar Partzsch, Department of Mathematics at Dresden University of Technology, for valuable comments and discussions.
\vspace{-0.5ex}


%* literature *****************************************************************
\bibliographystyle{IEEEtran}
\bibliography{./bibliography/glo2007}

\end{document}